\documentclass[%
reprint, 
superscriptaddress,
amsmath,
amssymb,
twocolumn,
]{revtex4-1}

\usepackage[normalem]{ulem}
\usepackage{graphicx}
\usepackage[english]{babel}
\usepackage[utf8]{inputenc}

\usepackage{bm}
\usepackage[]{upgreek}
\usepackage{gensymb}
\usepackage[]{nicefrac}
\usepackage{hyperref}
\usepackage{dsfont }
\usepackage[dvipsnames]{xcolor}
\usepackage{multirow}

\usepackage{natbib}

\usepackage{booktabs}
\AtBeginDocument{
\heavyrulewidth=.08em
\lightrulewidth=.05em
\cmidrulewidth=.03em
\belowrulesep=.65ex
\belowbottomsep=0pt
\aboverulesep=.4ex
\abovetopsep=0pt
\cmidrulesep=\doublerulesep
\cmidrulekern=.5em
\defaultaddspace=.5em
}

\newcommand{\beginsupplement}{%
        \setcounter{table}{0}
        \renewcommand{\thetable}{S\arabic{table}}%
        \setcounter{figure}{0}
        \renewcommand{\thefigure}{S\arabic{figure}}
        \setcounter{section}{0}
        \renewcommand{\thesection}{S\arabic{section}}
     }

\begin{document}

\title{Decoherence of dipolar spin ensembles in diamond}


\author{Erik Bauch}
\affiliation{Department	of	Physics,	Harvard	University,	Cambridge,	Massachusetts	02138,	USA}

\author{Swati Singh}
\affiliation{Electrical and Computer Engineering, University of Delaware, Newark, Delaware 19716, USA}

\author{Junghyun Lee}
\affiliation{Department of  Physics,  Massachusetts Institute of  Technology, Cambridge,  Massachusetts 02139,  USA}
\affiliation{Center for Quantum Information, Korea Institute of Science and Technology (KIST), Seoul, 02792, Republic of Korea}

\author{Connor A. Hart}
\affiliation{Department	of	Physics,	Harvard	University,	Cambridge,	Massachusetts	02138,	USA}

\author{Jennifer M. Schloss}
\affiliation{Department	of	Physics,	Massachusetts	Institute	of	Technology,	Cambridge,	Massachusetts	02139,	USA}
\affiliation{Center	for	Brain	Science,	Harvard	University,	Cambridge,	Massachusetts	02138,	USA}

\author{Matthew J. Turner}
\affiliation{Department	of	Physics,	Harvard	University,	Cambridge,	Massachusetts	02138,	USA}
\affiliation{Center	for	Brain	Science,	Harvard	University,	Cambridge,	Massachusetts	02138,	USA}

\author{John F. Barry}
\affiliation{Lincoln Laboratory, Massachusetts Institute of Technology, Lexington, Massachusetts 02420, United States}

\author{Linh Pham}
\affiliation{Lincoln Laboratory, Massachusetts Institute of Technology, Lexington, Massachusetts 02420, United States}

\author{Nir Bar-Gill}
\affiliation{The Hebrew University of Jerusalem, Jerusalem 91904, Israel}

\author{Susanne F.  Yelin}
\affiliation{Department	of	Physics,	Harvard	University,	Cambridge,	Massachusetts	02138,	USA}
\affiliation{Department of Physics, University of Connecticut, Storrs, Connecticut 06269, USA}

\author{Ronald L. Walsworth}
\email{rwalsworth@cfa.harvard.edu}
\affiliation{Department	of	Physics,	Harvard	University,	Cambridge,	Massachusetts	02138,	USA}
\affiliation{Center	for	Brain	Science,	Harvard	University,	Cambridge,	Massachusetts	02138,	USA}
\affiliation{Harvard-Smithsonian Center	for	Astrophysics,	Cambridge,	Massachusetts	02138,	USA}

\date{\today}

\begin{abstract} 

We present a combined theoretical and experimental study of solid-state spin decoherence in an electronic spin bath, focusing specifically on ensembles of nitrogen vacancy (NV) color centers in diamond and the associated substitutional nitrogen spin bath. We perform measurements of NV spin free induction decay times $T_2^*$ and spin-echo coherence times $T_2$ in 25 diamond samples with nitrogen concentrations [N] ranging from 0.01 to 300\,ppm. We introduce a microscopic model and perform numerical simulations to quantitatively explain the degradation of both $T_2^*$ and $T_2$ over four orders of magnitude in [N]. Our results resolve a long-standing discrepancy observed in NV $T_2$ experiments, enabling us to describe NV ensemble spin coherence  decay shapes as emerging consistently from the contribution of many individual NV.

\end{abstract}

\maketitle

Solid-state electronic spins have garnered increasing relevance as building blocks in a wide range of quantum science experiments~\cite{Saeedi2013,Hensen2015,Sipahigil2016a}. Recently, high-sensitivity quantum sensing experiments have been enabled by the NV spin ensemble system in diamond~\cite{Wolf2015, Barry2016, Chatzidrosos2017}. Such work exploits NV centers' millisecond-long spin lifetimes under ambient conditions~\cite{Doherty2013} and hinges on both the coherent microwave control and the optical initialization and readout of their spin states. In addition to enabling technological advances, these favorable properties make NV centers, and specifically NV ensembles, a leading platform for the study of novel quantum many-body physics and non-equilibrium spin dynamics~\cite{Choi2017a,Choi2017,Bauch2018}.

\begin{figure}[ht]
  \centering
  \includegraphics[]{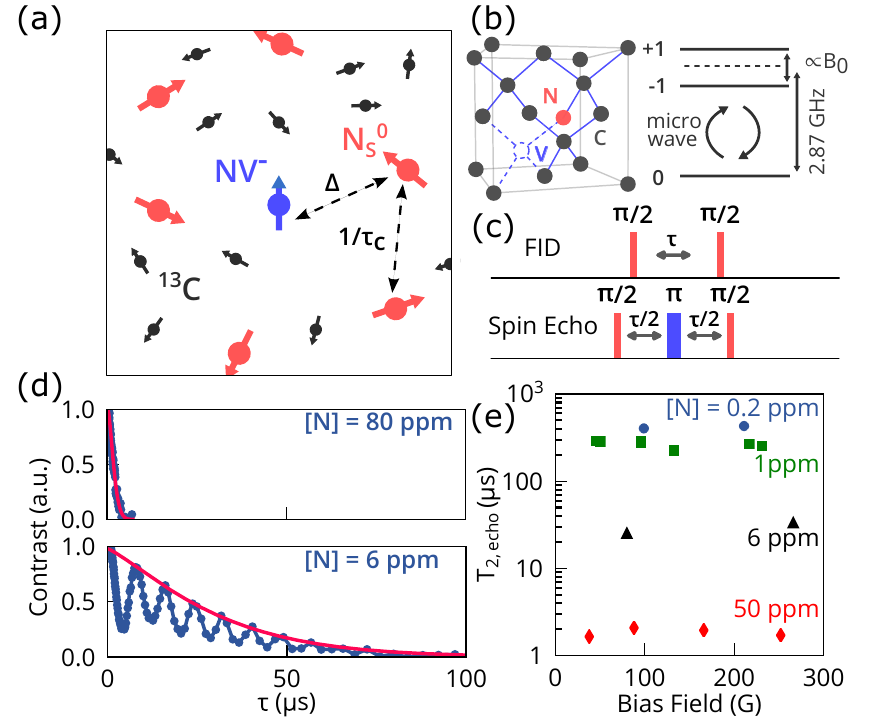}  
  \caption{
  a) Schematic of bath interaction in NV diamond showing NV$^\text{-}$ as central spin, substitutional nitrogen $\text{N}_\text{S}^0$ (P1 center, $S=1/2$) and $^{13}$C nuclear ($I=1/2$) bath spins. Central-spin-bath coupling strength $\Delta$ and intra-bath coupling strength $1/\tau_c$ are shown (see text for more information). 
  b) NV crystallographic defect and electronic spin-triplet ground-state level structure.
  c) Free induction decay (Ramsey) and spin echo pulse sequence are used to measure $T_2^*$ and $T_2$, respectively. 
d) Representative spin echo decay fringes (blue) and envelope (red) for [N] = 80 and 6\,ppm diamond samples. The modulation of the echo signal visible in $^{13}$C samples is caused by Larmor precession of the $^{13}$C nuclear spins (see main text and Suppl.\,\ref{app:measurement_details}).
  e) Magnetic field dependence of $T_2$ for select set of samples.
    }
  \label{fig:fig1}
\end{figure}

NV ensembles in diamond, like many solid-state spin systems, necessarily suffer from decay of electronic spin coherence (with characteristic times $T_2^*$ and $T_2$) and spin state population (with time $T_1$). Dipolar interactions within a complex spin bath environment may limit these relaxation times, bounding the achievable sensitivity of NV-ensemble-based quantum sensing devices and revealing rich many-body dynamics of dipolar-coupled spin systems (see Fig.\,\ref{fig:fig1}a)~\cite{Abragam1983}.

Solid-state spin-based sensing devices utilize host material widely ranging in concentrations of both electronic and nuclear spin species~\cite{Acosta2009, Balasubramanian2009, Wolf2015, Bauch2018}, which motivates investigation of dipolar-induced decoherence across varying concentrations of both like and unlike spin species. For example, in NV-rich samples, paramagnetic substitutional nitrogen impurities (P1 centers, $S=1/2$)~\cite{Smith1959,Cook1966,Loubser1978} typically persist at concentrations similar to or exceeding the NV concentration, setting the NV spin relaxation time scales. In other dense spin-ensemble systems, spin relaxation has been observed to be dominated by like-spin interactions~\cite{VanWyk1997,Tyryshkin2012,Choi2017a}. Understanding the degree and character of decoherence informs engineering of material and design of spin interrogation schemes for high-performance quantum devices.

In this Letter, we experimentally and theoretically investigate the NV ensemble electronic spin dephasing time $T_2^*$ and spin echo coherence time $T_2$ in diamond samples with nitrogen density spanning more than four orders of magnitude. We purposefully choose diamond samples with grown-in spin concentrations $[\text{NV}] \ll [\text N]$ to limit the study to NV decoherence resulting from paramagnetic nitrogen bath spins. In this regime, we identify characteristic inverse-linear scalings for both NV $T_2^*$ and $T_2$ with nitrogen density $[\text N]$. We also find NV ensemble spin coherence decay shapes that can be consistently described as the result of ensemble averaging. Our work extends the comprehensive knowledge acquired for single NVs~\cite{Balasubramanian2009,Mizuochi2009,DeLange2010,DeLange2012} to NV ensembles, and thus is critical for the development of ensemble-based quantum applications.


{\it Experimental Results --} Our study comprises of 20 natural abundance diamond samples ($[^{13}\mathrm{C}] = 1.1\,\%$, ``$^{13}$C-samples") and 5 isotopically enriched samples ($[^{13}\mathrm{C}] \lesssim 0.05\,\%$, ``$^{12}$C-samples") with total nitrogen concentrations in the range [N] = 10\,ppb - 300\,ppm. Free induction decay (FID) NV $T_2^*$ and spin echo $T_{2}$ measurements are performed (see Fig.\,\ref{fig:fig1}c) using confocal or wide field microscopy. In both experimental setups, 532\,nm laser light is applied to optically initialize and readout the NV spin polarization. In addition, we apply a static magnetic field $B_0$ along one of the [111] crystal directions (misalignment angle $\leq 3\degree$), which singles out one of the four possible NV orientations and lifts the $|\pm 1\rangle$ degeneracy of the NV spin-1 ground state (see Fig.~\ref{fig:fig1}b). Pulsed microwaves resonant with the $|0\rangle\leftrightarrow |-\!1\rangle$ or $|0\rangle\leftrightarrow |+\!1\rangle$ spin transition are deployed to coherently manipulate the NV spin state.


In Fig. \ref{fig:fig1}d we depict two spin echo curves from diamonds with [N] = 80 and 6\,ppm, which are representative of the set of 20 $^{13}$C diamond samples studied in this work. The data exhibit coherent modulation of the NV spin echo signal due to Larmor precession of nuclear bath spins (see Suppl.~\ref{app:measurement_details})~\cite{Rowan1965,Childress2006,Stanwix2010}. Isotopically enriched $^{12}$C-samples did not exhibit any modulation of the coherence signal irrespective of the applied magnetic field strength. The focus of this work is the overall exponential-type decay, which is associated with loss of NV ensemble electronic spin coherence due to dipolar interactions with electronic nitrogen bath spins~\cite{DeLange2010}. For all samples, the decay envelope was subsequently fitted to the form $C_0 \exp{[-(t/T_2)^p]}$ (red solid lines in Fig.\,\ref{fig:fig1}d) to extract $T_{2}([\text N])$ and the stretched exponential parameter $p$~\cite{DeSousa2009,DeLange2010}.

\begin{figure}[ht]
  \centering
  \includegraphics[]{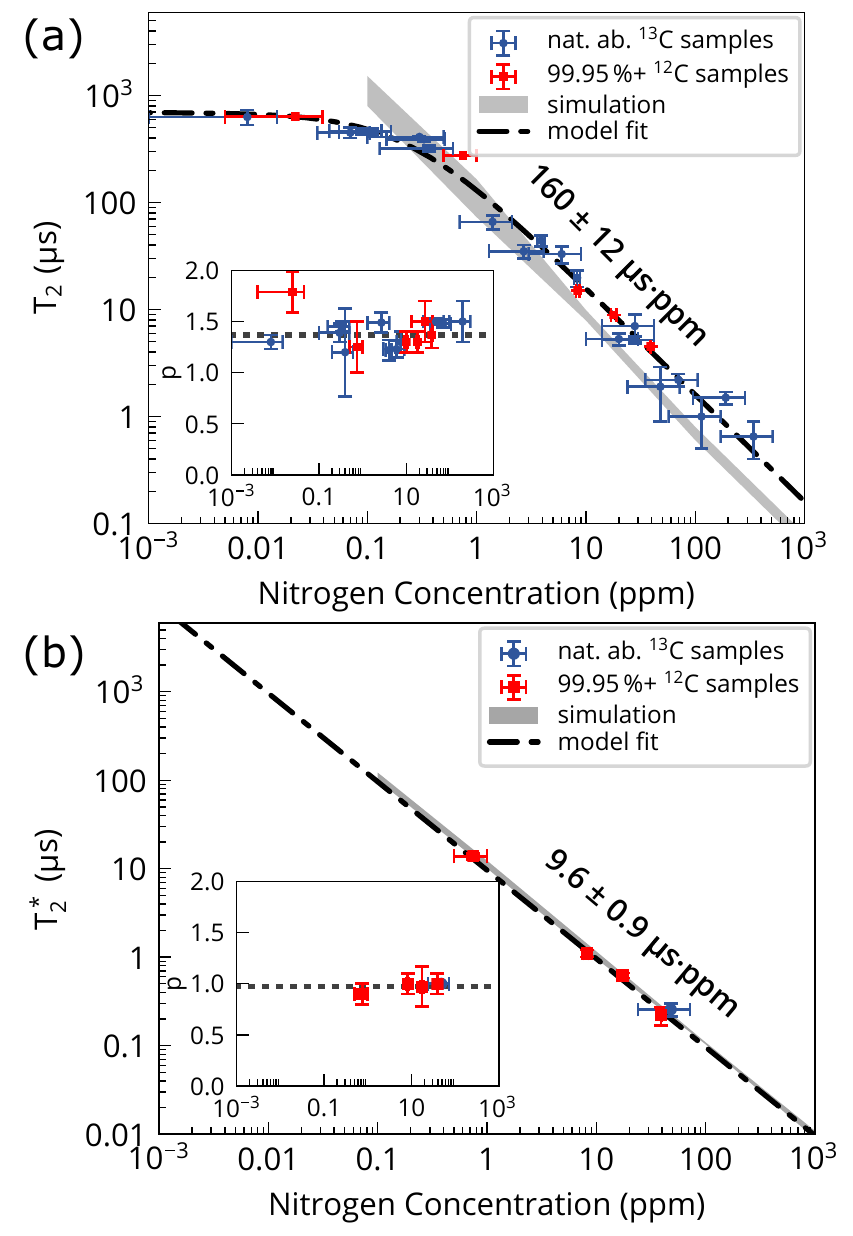}  
  \caption{
  NV ensemble coherence
  a) NV ensemble spin echo $T_2$ as a function of [N] with orthogonal-distance regression fit to Eqn.\,\ref{eqn:eq1} (black dashed line) and range of values extracted from numerical simulation (gray band). Uncertainties are $95\,\%$ confidence intervals. Inset: Stretched exponential parameters $p$ extracted from measurements. The average value is indicated as black dotted line. b) Similarly, NV ensemble $T_2^*$ as a function of nitrogen concentration [N].
    }
  \label{fig:fig2}
\end{figure}

Figure\,\ref{fig:fig1}e shows $T_{2}$ times derived from this analysis for a select set of $^{12}$C and $^{13}$C samples as a function of magnetic bias field strengths. Only small variations in $T_2$ ($\lesssim 10\,\%$) are observable for the range of bias field strengths $B_0$ ($2 - 30\,$mT, see Suppl.\,\ref{app:measurement_details}) indicating that $T_2$ is largely independent of magnetic field.
Next, we summarize in Fig.\,\ref{fig:fig2}a $T_{2}$ values for all samples as a function of total nitrogen concentration $[\text N]$ with two regimes discernible. At nitrogen concentrations $[\text N] \gtrsim 0.5\,$ppm, $T_{2}$ exhibits an inverse-linear dependence on the nitrogen concentration, suggesting that interactions with nitrogen bath spins are the dominant source of decoherence of NV ensemble electronic spin. This inverse-linear scaling is consistent with studies in comparable crystalline systems and characteristic of dipolar-coupled spin environments~\cite{VanWyk1997,Mizuochi2009,Abe2010,Stepanov2016}. In such instances, $1/T_2$ is proportional to the spin bath density $n_\text{bath}$, i.e., $1/T_2 = B\,\cdot\,n_{bath}$, where the factor $B$ depends on microscopic details of the system-bath coupling and the spin bath dynamics (see below).

For samples with concentrations $[\text N] \lesssim 0.5$\,ppm, however, $T_2$ saturates at $\approx 700\,\upmu$s, for both isotopically purified and natural abundance samples. This $T_2$ bound is well below the observed limit set by NV electronic spin lattice relaxation, where previous work found that $T_{2,\text{max}} \approx T_1/2 \approx 2.5\,$ms~\cite{Bar-Gill2013}. While a detailed investigation of the measured $T_2$ bound is beyond the scope of this manuscript, additional decoherence mechanisms may include the $^{13}$C nuclear spin bath (for $^{13}$C samples)~\cite{Mizuochi2009, Hall2014}, spurious external AC magnetic fields, and other spin defects in the diamond host. 
%

To further quantify the observed scaling, we fit the extracted $T_2$ values to the form
\begin{equation} \label{eqn:eq1}
    1/T_2([\text N]) = B_\text{NV-N} \cdot [\text N] + 1/T_{2,\text{other}},
\end{equation}
where $B_\text{NV-N}$ is the nitrogen-dominated NV decoherence rate per unit density and $T_{2,\text{other}}$ accounts for decoherence mechanisms independent of nitrogen (including spin-lattice relaxation~\cite{Bar-Gill2013}). From the fit (black dashed line in Fig.\,\ref{fig:fig2}a) we extract $B_\text{NV-N} = 2\pi \times (1.0\pm 0.1)\,$kHz/ppm $(1/B_\text{NV-N} = 160 \pm 12\,\upmu$s$\,\cdot\,$ppm) and $T_{2,\text{other}} = 694 \pm 82\,\upmu$s. We also fit this model to the natural abundance $^{13}$C (blue dots) and $^{12}$C (red squares) sample data alone, and find agreement within error margins among the three sets of diamond samples (see Suppl.~\ref{app:measurement_details}). Additionally, we plot the extracted stretched exponential parameter $p$ in the inset of Fig.\,\ref{fig:fig2}a. All samples exhibit exponential-type decay with a sample average value $\overline p = 1.37 \pm 0.23$. This non-integer spin-echo decay is in striking contrast to the cubic decay ($p=3$) observed for single NV centers in a bath of nitrogen spins~\cite{DeSousa2009,DeLange2010} (see discussion below).


Similarly, FID Ramsey measurements were employed to determine the NV ensemble electronic spin dephasing time $T_2^*([\text N])$. A recent related study~\cite{Bauch2018} demonstrated that several dephasing mechanisms typically limit $T_2^* \lesssim 1 \upmu$s in NV ensemble samples. Sources of ensemble dephasing include interactions with nuclear $^{13}$C bath spins~\cite{Mizuochi2009,Dreau2012,Bauch2018}, crystal-lattice strain fields over the diamond~\cite{Jamonneau2015,Bauch2018}, and measurement-related artifacts such as magnetic field gradients over the collection volume and temperature fluctuations~\cite{Acosta2010,Bauch2018}. Great care was therefore taken to isolate the nitrogen-specific contribution to $T_2^*$ from other contributions. In particular, we limit our set of FID measurements to $^{12}$C samples in the $[\text N] = 1 - 100$\,ppm range for which $^{13}$C-related dephasing can be neglected~\cite{Bauch2018}. Moreover, we sense dephasing in the NV double quantum basis $\{+1,-1\}$~\cite{Fang2013,Mamin2014,Bauch2018} to mitigate contributions from strain field gradients and temperature fluctuations of the diamond sample. We correct for the twice higher dephasing rate in the double quantum basis; further details are provided in Suppl.\,\ref{app:measurement_details} and Ref.~\cite{Bauch2018}.)

NV ensemble electronic spin dephasing times $T_2^*$ and stretched exponential parameters $p$ were extracted from the FID data via a fit of the decay envelopes to the form $C_0 \exp{(-t/T_2^*)^p}$. Figure~\ref{fig:fig2}b shows the measured $T_2^*$ and $p$ values as a function of [N] for the subset of samples. Similarly to $T_{2}$ (Fig.\,\ref{fig:fig2}a), we find $T_2^*$ to scale inverse-linear with nitrogen concentration in the explored regime. We consequently fit the data to the form $1/T_2^*([\text N]) = A_\text{NV-N} \cdot [\text N]$ (compare to Eqn.\,\ref{eqn:eq1}), where $A_\text{NV-N}$ is the nitrogen-related NV ensemble dephasing rate per unit density. From the fit we extract $A_\text{NV-N} = 2\pi \times (16 \pm 1.5)\,$kHz$\cdot$ppm $(1/A_\text{NV-N} = 9.6 \pm 0.9\,\upmu$s$\,\cdot\,$ppm). In addition, all FID NV ensemble measurements exhibit simple exponential decay ($p=1$, see Fig.\,\ref{fig:fig2}b inset), which deviates from the quadratic decay $(p=2)$ observed for single NV measurements~\cite{DeLange2010, DeLange2012}, and is in agreement with earlier theoretical work~\cite{Dobrovitski2009}.


{\it Theoretical Model --} To support the assumption that the measured scaling for $T_{2}([\text N])$ and $T_2^*([\text N])$ are indeed due to nitrogen bath spins, we developed an analytic model and performed numerical simulations of the bath dynamics. We started with a phenomenological model for incoherent spin bath dynamics that leads to dephasing and decoherence of NV center electronic spins. Under the secular approximation, the dipolar interaction between two spins $i$ and $j$ can be simplified to \cite{Abragam1983}
\begin{equation}
H_{ij}=C_{\parallel}^{ij} S_z^i S_z^j +C_{\bot}^{ij} (S_+^i S_-^j + S_-^i S_+^j).
\label{eq:Hij1}
\end{equation}
Transforming the dipole-dipole interaction to a frame along the [111] crystal axis, the two coefficients are
\begin{eqnarray}
C_{\parallel}^{ij}&=&\frac{\mu_0}{4\pi}\frac{\hbar^2 \mu_e^2}{r_{ij}^3} (1-2 \cos^2{\theta})\text{, and} \\
C_{\bot}^{ij}&=&\frac{\mu_0}{4\pi} \frac{\hbar^2 \mu_e^2}{2 r_{ij}^3} \left( 1-\frac{1}{4} \sin^2{\theta} \right).
\label{eq:DDcoefficients}
\end{eqnarray}
Here, $r_{ij}$ is the distance, and $\theta$ is the polar angle between the applied bias magnetic field and the intra-spin axis (i.e., a vector connecting the two spins). For the case of dipolar interactions between NV and a P1 bath spin, the large NV zero field splitting makes spin flip-flop transitions non-energy-conserving and the second term in Eqn.\,\ref{eq:Hij1} is strongly suppressed. The system-bath interaction thus simplifies to $H_\text{NV-N} = \sum_i C_{\parallel}^{\text{NV},i}S_z^\text{NV} S_z^i$ and the spin bath may be treated as an effective B-field whose value depends on the state of the bath spins. At room temperature, all spin states are equally occupied, and we can define the system-bath interaction strength $\Delta^2 \equiv \sum_i (C_{\parallel}^{\text{NV},i})^2/4$, to characterize the effective B-field due to dipolar coupling between the NV sensor spin and N-bath spins. $\Delta$ quantifies the broadening of the NV sensor electron spin resonance and dephasing rate of the NV qubit.

\begin{figure}[ht]
\begin{center}
\includegraphics[width=3in]{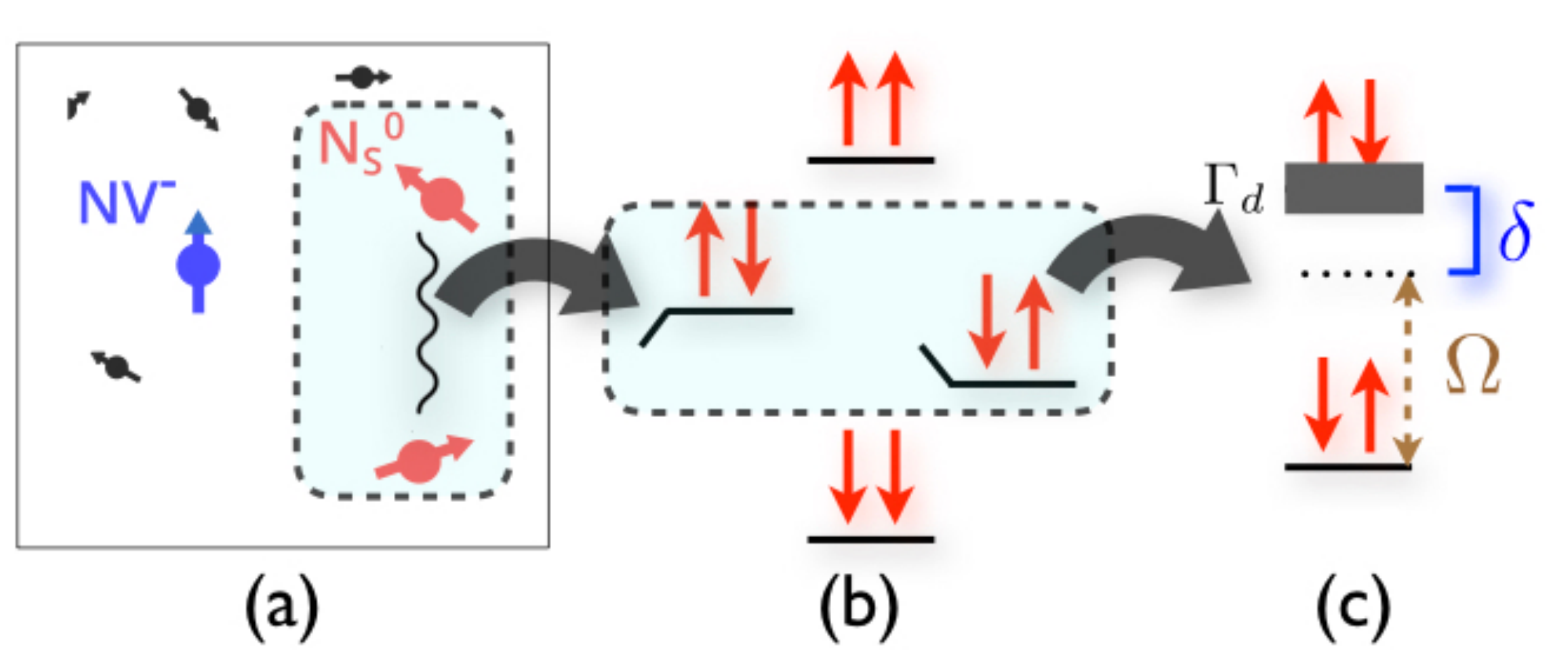}
\caption{Phenomenological model to evaluate bath correlation time $\tau_c$. (a) The bath spin dynamics arise due to interaction between nearby P1 spins. (b) Out of the four configurations, states $|\uparrow\downarrow\rangle$ and $|\downarrow\uparrow\rangle$ give spin bath dynamics under Hamiltonian in Eqn.\,\ref{eq:Hij1}. (c) These dynamics can be mapped to a spin 1/2 system. The slow dynamics of the remaining spin bath is taken into account as a dephasing term in the master equation.}
\label{fig:dephasing}
\end{center}
\end{figure}

{We now extent our model to evaluate the rate of P1 spin bath dynamics.} In the presence of an external B-field, the flip-flop interaction between two P1 spins in Eqn.\,\ref{eq:Hij1} can be mapped to a pseudo-spin-$1/2$ system, with Hamiltonian $H_\text{ps}= \hbar \delta \sigma_z + \hbar \Omega \sigma_x $, where $\delta$ is the difference in the local field (including Overhauser) experienced by the two bath spins and $\Omega=C_{\bot}^{ij}$. {As shown in Fig.\,\ref{fig:dephasing}, the pseudo-spin states $|e,g\rangle$ correspond to $|e\rangle = |\!\uparrow\downarrow\rangle$ and $|g\rangle = |\!\downarrow\uparrow\rangle$ states of the P1 spin pair.}
In numerically evaluating $\delta$ for simulations, we took into account the P1 nuclear spin state ($I=1$), the P1 Jahn-Teller axis (along any of the four [111] diamond crystal axes)~\cite{Ammerlaan1981}, and nearby P1 centers. Along with the coherent Hamiltonian evolution, there is a stochastic change in $\delta$ due to changes in the local spin environment caused by (among other things) flip-flop dynamics of far away bath spins. Including this incoherent part, the evolution of the density matrix is given by $\dot{\rho}=-i/\hbar[H_\text{ps},\rho]+\mathcal{L}[\rho]$, where the Louvillian includes dephasing at rate $\Gamma_{d}$. {Here $\Gamma_d$ denotes the rate of change of the effective B-field to which the pseudospin is subjected.}

We solve for the rate of change of population ($R_{\rm flip}^{ij}$) to obtain
\begin{equation}
R_{\rm flip}^{ij}=\frac{\Omega^2}{\Gamma_d}\frac{\Gamma_d^2}{\Gamma_d^2+\delta^2}.
\label{eq:Rflip}
\end{equation}
Assuming all P1 spin pairs act independently, we add the different rates in order to get an overall rate of spin bath dynamics $R_{\rm Tot} = \sum_{\{i,j\}} R_{\rm flip}^{ij}$. Realistically, each spin pair has a different $\delta$ and $\Gamma_d$. Nonetheless, while we evaluated $\delta$ for each P1 spin numerically, we assumed all pairs have the same dephasing rate $\Gamma_d\approx\sqrt{N_b}\bar{C}_{\parallel}^{ij}$, where $N_b$ is the number of bath spins and $\bar{C}_{\parallel}^{ij}$ is the average dipolar interaction. Note that $\Gamma_d$ is the intrinsic linewidth of the dipolar spin bath~\cite{Stamps2000}. We define $\tau_c\equiv 1/R_{\rm Tot}$ to characterize the spin bath correlation time~\cite{DeSousa2009,DeLange2010}. The many-body dynamics of the spin bath leads to decoherence of the NV center.

While the origin of $\Delta$ and $\tau_c$ are quantum in nature, in order to estimate the electronic spin coherence properties of NV ensembles, we can model the P1 bath dynamics as an effective random B-field with root-mean-square strength $\Delta$ and correlation time $\tau_c$~\cite{DeSousa2009,DeLange2010}. This approximation works well due to the low density of impurity spins and slow bath dynamics, allowing us to ignore higher order quantum effects. In our analytic model for the NV spin coherence signal, the effect of the nitrogen bath is modeled as an  Orenstein-Uhlenbeck stochastic process \cite{Klauder1962,DeSousa2009,Witzel2014}. While the details of the derivation are provided in the Suppl., here we present the main results. 

For a single NV initially polarized in the $|0\rangle$ state and undergoing the $|0\rangle \leftrightarrow |-1\rangle$ (or $|0\rangle \leftrightarrow |+1\rangle$) transition, the probability to be measured in state $|0\rangle$ after time $t \ll \tau_c$ is given by $p_\text{FID}^\text{single}(t) =\frac{1}{2}\left[ 1+e^{-(t/T_{2,\text{single}}^*)^2}\right]$ and $p_\text{echo}^\text{single}(t) = \frac{1}{2}\left[ 1+e^{-(t/T_{2,\text{single}})^3}\right]$~\cite{DeSousa2009,DeLange2010} (see Suppl.\,\ref{treatment_of_spinbath}) for FID and spin echo experiments, respectively. Here, $T_{2,\text{single}}^*=\sqrt{2}/\Delta_\text{single}$ and $T_{2,\text{single}}=(12 \tau_\text{c,single}/\Delta_\text{single}^2)^{1/3}$, where (as discussed previously) $\Delta_\text{single}$ characterizes the system-bath coupling strength and $\tau_\text{c,single}$ characterizes the spin bath correlation time. In order to evaluate NV ensemble averaged probabilities, we need to integrate $p_\text{FID}^\text{single}$ and $p_\text{echo}^\text{single}$ over the distribution of $\Delta_\text{single}$ and  $\tau_\text{c,single}$ in the ensemble. 
\begin{figure}[ht]
  \centering
  \includegraphics[]{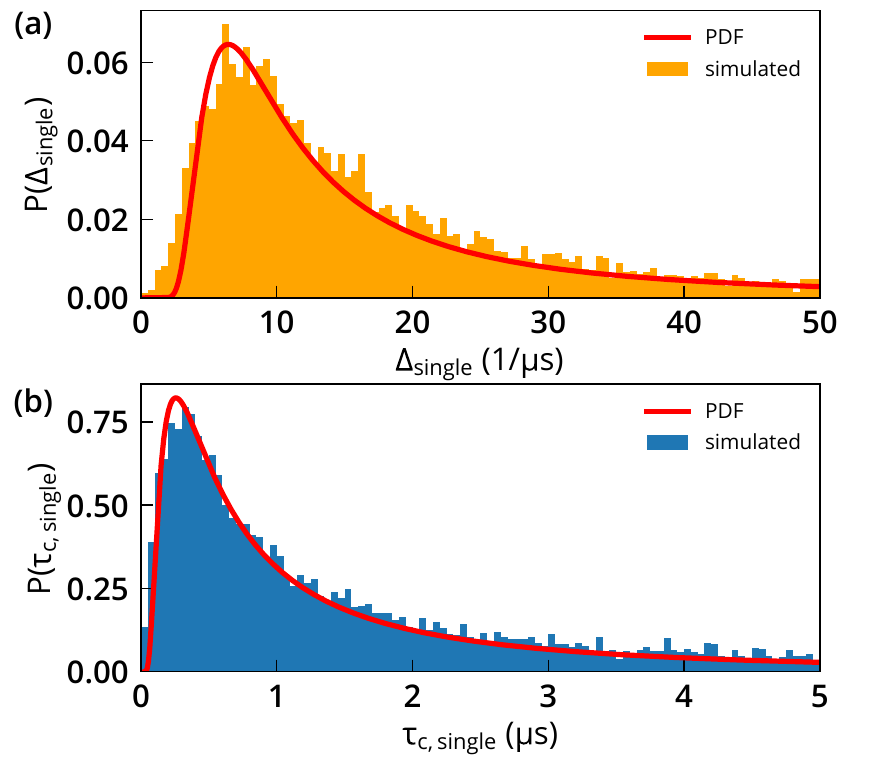} 
  \caption{Simulated distribution of coupling strength $\Delta$ (a) and bath correlation time $\tau_c$ (b) for $[\text N]=100 \,$ppm extracted from $10^4$ spin bath configurations. From a fit of each distributions (red line) to the analytic PDF P($\Delta$) and P($\tau_c$) given in the main text, respectively, the relevant ensemble parameters $\Delta_\text{e}$ and $\tau_\text{c,e}$ are extracted, and $T_2^*$ and $T_{2}$ calculated. Similar distributions are obtained for $[\text N]= 0.1, 1, 10$, and 1000\,ppm.
    }
  \label{fig:fig4}
\end{figure}

Figure\,\ref{fig:fig4} shows the distribution of $\Delta_\text{single}$ and $\tau_\text{c,single}$ generated from simulating $10^4$ bath configurations for $[\text N] = 100\,$ppm. For dipolar electronic spin interactions, the probability density function (PDF) for $\Delta_\text{single}$ has the form
$P(\Delta_\text{single})= \frac{\Delta_\text{ens}}{\Delta_\text{single}^2}\sqrt{\frac{2}{\pi}} e^{-\Delta_\text{ens}^2/2\Delta_\text{single}^2}$~\cite{Dobrovitski2008}, %
where $\Delta_\text{ens}$ is the average electronic spin-spin coupling strength within the NV ensemble. Integrating $p_\text{FID}^\text{single}(t)$ over the distribution of $\Delta_\text{single}$, we arrive at the ensemble averaged probability,
\begin{equation}\label{eq:FIDensm1}
p_\text{FID}^\text{ens}(t) =\frac{1}{2}\left[ 1+e^{-(t/T_{2,\text{ens}}^*)}\right],
\end{equation}
where $T_{2,\text{ens}}^*=1/\Delta_\text{ens}$. The ensemble-averaged NV FID signal thus exhibits simple exponential decay ($p=1$) in agreement with our experimental results (see Fig.\,\ref{fig:fig2}b inset). 

A similar analysis can be done to evaluate the NV ensemble $T_{2}$. In this case, we also need to account for the distribution of $\tau_\text{c,single}$. In our classical treatment, $\tau_c$ can be interpreted as a first passage time for a stochastic process (here several bath spin flip-flops) with PDF $P(\tau_\text{c,single})=\sqrt{\lambda/(2 \pi \tau_\text{c,single}^3)} e^{-\lambda(\tau_\text{c,single}-\tau_\text{c,ens})^2\big/2\tau_\text{c,single}\tau_\text{c,ens}^2}$~\cite{Grimmett2001}. Here, $\lambda$ and $\tau_\text{c,ens}$ are the shape parameter and the ensemble mean of the distribution, respectively. As shown in Fig.\,\ref{fig:fig4}, the above PDFs model the numerical distribution of $\Delta_\text{single}$ and $\tau_\text{c,single}$ obtained from simulating the dipolar-coupled system. Finally, we integrate over the PDFs of $\Delta_\text{single}$ and $\tau_\text{c,single}$ and obtain
\begin{equation}\label{eq:SEensm1}
p_\text{echo}^\text{ens}(t) \approx\frac{1}{2}\left[ 1+e^{-(t/T_{2,\text{ens}})^{3/2}}\right],
\end{equation}
where $T_{2,\text{ens}}=(2\tau_\text{c,ens}/\Delta_\text{ens}^2)^{1/3}$ ($t \ll \tau_\text{c,ens}$), and the ensemble-averaged spin echo decay exhibits non-integer decay with $p \approx 3/2$. This analytic treatment agrees with the experimentally measured form of the NV ensemble spin echo decay envelopes (see Fig.\,\ref{fig:fig2}a inset). Furthermore, using the bath simulation results for $\Delta_\text{ens}$ and $\tau_\text{c,ens}$ from our phenomenological model in Eqns.\,\ref{eq:FIDensm1} and \ref{eq:SEensm1}, we find good agreement with experiments for $T_2^*([\text N])$ and reasonable agreement for $T_2([\text N])$, as shown in Fig.\,\ref{fig:fig2}. The consensus among analytic results, experiment and numerical simulation leads us to conclude that the Orenstein-Uhlenbeck model adequately describes the decoherence of an ensemble of NV electronic spins induced by nitrogen bath spins, with $A_\text{NV-N}$ and $B_\text{NV-N}$ as the experimentally determined dephasing and decoherence rates (per unit density) due to nitrogen bath spins. These results are consistent with earlier theoretical predictions that similar electronic spin baths can be modeled classically~\cite{Witzel2014}. We emphasize that the role of the $^{13}$C bath is negligible if nitrogen is the dominant source of NV ensemble electronic spin decoherence. 

Taking the ratio $T_{2}([\text N])/T_{2}^*([\text{N}]) = A_\text{NV-N}/B_\text{NV-N}$, which is independent of absolute nitrogen concentration, we find that $T_{2}$ exceeds $T_2^*$ by $\sim 16\times$ across a wide range of diamond samples and [N]. Given the provided scalings, rough calibration of bulk substitutional nitrogen spin concentrations through NV coherence measurements can thus be performed.

Finally, we comment on the $p$ parameter in single and ensemble NV experiments. For single NVs, the exponential-type decay shapes for FID, spin echo, and spin lifetime measurements are determined by spin-resonance experiments in nitrogen-rich diamonds, showing good agreement with theory (see Tab.\,\ref{tab:tab1}, columns 2 and 3).

\begin{center}
\begin{table}[ht!]
    \begin{tabular}{ccccc} \toprule
    \multirow{2}{*}{Experiment} & \multicolumn{2}{c}{single NV}  & \multicolumn{2}{c}{NV ensemble}\\
    \cmidrule{2-3} \cmidrule{4-5} 
    & $p$ & exp./theo. & $p$ & exp./theo.\\
    \midrule
    $T_{2}$& 3 &~\cite{DeLange2010}/\cite{DeSousa2009} & $\approx 3/2$ & this work \\
    $T_2^*$ & 2 &~\cite{DeLange2010} /\cite{DeSousa2009} & 1     &~\cite{Bauch2018} /~\cite{Dobrovitski2009}, this work\\
     $T_1$ (cross-relax.) & -   & - & 1/2 &~\cite{Jarmola2015,Choi2017a} /~\cite{Hall2016,Choi2017a} \\
    $T_1$ (spin-lattice relax.) & 1     & \cite{Rosskopf2014InvestigationDiamond} & 1     &~\cite{Redman1991,Jarmola2012}\\
    \bottomrule
    \end{tabular}%
\caption{Stretched exponential parameter $p$ for single NV and NV ensemble measurements.}
 \label{tab:tab1}%
\end{table}%
\end{center}

However, as has been observed in several experiments, different power laws emerge for the decay signal of individual NV centers with slightly varying spin environments. For example, several studies  independently reported a square-root decay shape ($p=1/2$) for NV $T_1$ spin relaxation in high-NV density samples~\cite{Jarmola2015,Hall2016,Choi2017a}. In these samples, the NV $T_1$ is limited by cross-relaxation within the strongly interacting bath of NV$^\text{-}$ spins rather than spin-lattice relaxation (i.e., phononic decay), which exhibits a simple exponential decay envelope~\cite{Redman1991,Takahashi2008,Jarmola2012}. This situation is comparable the present experiment, for which the measured NV ensemble $T_2^*$ and $T_{2}$ are the sum of many individual decays, with decay rates limited by dipolar interactions with an inhomogenous nitrogen spin bath. We stress that the form of the stretched exponential from such NV ensembles results from the averaging of NV and spin bath distributions, which in turn depends on the microscopic quantum interactions that lead to decay. Unlike $T_2^*$, the NV ensemble $T_2$ depends on the many-body dynamics of the spin bath. Mean field models like the one used here are insufficient to accurately capture the dynamics of such highly disordered quantum systems. While our model describes the relevant scalings and NV coherence decay shapes, the underestimation of $T_2$, particularly for larger NV and nitrogen concentration, may indicate localization-type behavior. Thus future work should include theoretical and experimental investigation of localized phases in this system.

In conclusion, we studied electronic spin dephasing and decoherence of NV ensembles in diamond as a function of nitrogen concentration and extracted the scalings for $T_{2}^*([\text N])$ and $T_{2}\text{[N]}$. We found that the dominant nitrogen electronic spin bath dynamics are well described by an Orenstein-Uhlenbeck process, with reasonable agreement between experimental measurements and numerical simulations. We also analyzed the NV ensemble spin coherence decay shape for FID and spin echo measurements, and found them well explained by a statistical average over the contribution of many individual NV signals, in analogy to the decay shapes observed in NV $T_1$ ensemble experiments.


We acknowledge fruitful discussion with Joonhee Choi and Soonwon Choi. We thank EAG Laboratories for the SIMS measurements of nitrogen concentration. This material is based upon work supported by, or in part by, the United States Army Research Laboratory and the United States Army Research Office under Grant No. W911NF1510548; the National Science Foundation Electronics, Photonics and Magnetic Devices (EPMD), Physics of Living Systems (PoLS), and Integrated NSF Support Promoting Interdisciplinary Research and Education (INSPIRE) programs under Grants No. ECCS-1408075, PHY-1504610,  and EAR-1647504, respectively; and Lockheed Martin under award A32198. This work was performed in part at the Center for Nanoscale Systems (CNS), a member of the National Nanotechnology Coordinated Infrastructure Network (NNCI), which is supported by the National Science Foundation under NSF award no. 1541959. J. M. S. was supported by a Fannie and John Hertz Foundation Graduate Fellowship and a National Science Foundation (NSF) Graduate Research Fellowship under Grant 1122374.

\widetext
\clearpage
\begin{center}
\textbf{\large Supplemental Materials: Decoherence of dipolar spin ensembles in diamond}
\end{center}

\beginsupplement

\section{Sample Characterization}
 
All diamond crystals were manufactured by Element Six and Apollo Diamond~\footnote{now Scio Diamond Technology Corporation}. Samples with nitrogen spin densities [N]$\lesssim 100\,$ppm were grown using chemical vapor deposition (CVD, for a review, see~\cite{Schwander2011}) and consist of bulk diamond plates, as well as thin ($\lesssim 100\,\upmu$m) nitrogen-doped layers grown on top of Ib or IIa diamond substrates. Two other samples were grown using the high-pressure-high-temperature (HPHT) method~\cite{Bundy1955,Sumiya1996,Kanda2000,Hartland2014,Zaitsev2001} and cover the range $[\text N] \gtrsim 100\,$ppm. Due to the low abundance of NV center spins in all samples ($[\text{NV}] \ll [\text{N}]$), contributions from NV-NV dipolar interactions are negligible for the experiments in this manuscript.

The total nitrogen concentration within diamond samples studied in this work is determined by several methods: For the majority of samples, the manufacturers provided estimated nitrogen spin concentrations or secondary ion mass spectroscopy (SIMS) data taken on samples from the same growth run. In addition, for a subset of samples SIMS measurement were performed by EAG Laboratories. In select samples, nitrogen was determined through fourier-transformed infrared spectroscopy (FTIR)~\cite{Taylor1990,Hartland2014}. 

The reported uncertainties in [N] are calculated from the mean and variation in spin concentration values provided by combining results from various methods. If only one method was available, the uncertainty is given by the method's reported error margin. For SIMS measurements an error of $50\,\%$ is conservatively assumed, which also accounts for variations in N throughout different parts of a sample.

\section{Measurement Details}\label{app:measurement_details}

\textit{Spin Echo --} Low nitrogen density $^{13}$C samples exhibited periodic modulation of the NV spin echo signal (ESEEM, Ref.~\cite{Rowan1965,Childress2006}) owing to the Larmor precession of the $^{13}$C nuclear spin bath, as shown in Fig.\,\ref{fig:fig1}d ($[\text{N}] = 6\,$ppm sample). Revival and collapses of the spin echo signal occurred with frequency $f_{Larmor} = \frac{\gamma_{^{13}\text{C}}}{2\pi}B_0$, where $\gamma_{^{13}\text{C}} = 2\pi \times 1.07\,$MHz/T~\cite{Childress2006,Stanwix2010} is 
the $^{13}$C nuclear gyromagnetic ratio and $B_0$ is the bias magnetic field. To clearly separate the overall decay envelope from the Larmor signal, for each natural abundance $^{13}$C sample the bias field was adjusted between $2 - 30\,$mT to tune the Larmor precession frequency such that $f_{Larmor} \gg 1/T_2$ (low nitrogen samples) or $f_{Larmor} \ll 1/T_2$ (high nitrogen samples). The isotopically enriched $^{12}$C samples did not exhibit modulation of the coherence signal independent of the applied magnetic field strength. As discussed in the main text and Fig.\,\ref{fig:fig1}e, only small variations in NV $T_2$ are observed with changing $B_0$, which suggests that $T_2$ is largely independent of bias magnetic field for the range of fields. 

We also fit Eqn.\,\ref{eqn:eq1} to the $^{13}$C and $^{12}$C sample NV $T_2$ data alone (see Fig.\,\ref{fig:fig2}a) and obtained \{$A_\text{NV-N} = 2\pi \times (1.0 \pm 0.2)\,$kHz/ppm, $T_{2,\text{other}} = 715 \pm 248\,\upmu$s\} and \{$A_\text{NV-N} = 2\pi \times (0.9 \pm 0.1)\,$kHz/ppm, $T_{2,\text{other}} = 657 \pm 94\,\upmu$s\}, respectively. The close agreement (within error margins) among $A_\text{NV-N}$ values extracted for  $^{13}$C,  $^{12}$C, and the combined data suggests that NV spin ensemble decoherence due to $^{13}$C nuclear spins in natural abundance samples on $T_2$ is negligible, when nitrogen is the dominant source of decoherence.

\textit{Ramsey --} Several inhomogeneous broadening mechanisms contribute to NV ensemble spin dephasing. For example, the $T_2^*$-limit in natural abundance samples set by $1.1\% ^{13}$C spins ($\sim 1~\upmu$s~\cite{Bauch2018}) restricts our measurements to isotopically enriched $^{12}$C samples. In addition, FID decay was probed in the NV center's double quantum basis ($\{-1, +1\}$) to mitigate effects of strain fields and temperature fluctuations. To account for the twice higher gyromagnetic ratio and doubled dephasing rate in the double quantum basis, the extracted $T_2^*$ values are multiplied by 2. Lastly, measurements where performed at low magnetic bias fields ($\lesssim 20\,$G) to reduce the influence of magnetic field gradients. Further experimental details are given in Ref.~\cite{Bauch2018}.

\section{Classical treatment of the spin bath dynamics}\label{treatment_of_spinbath}

In this section, we discuss the evolution of several standard pulse sequences on an NV electronic spin in the presence of a fluctuating B-field. We also perform ensemble averages over a distribution of NV centers in order to evaluate the form of experimentally measured signals.

For an NV center undergoing dynamical decoupling pulse sequences under the presence of a fluctuating B-field, the probability to find the NV electronic spin in state $|0\rangle$ after time $\tau$ is given by
\begin{equation}
p(\tau)=\frac{1}{2}\left( 1+\langle \cos{\phi(\tau)}\rangle \right) = \frac{1}{2}\left( 1+\Re [\langle e^{i\phi(\tau)}\rangle ] \right).
\label{eq:ProbrandomB}
\end{equation}
Here, the total phase difference accumulated is
\begin{equation}
\phi(t)=\frac{g\mu_B}{\hbar}\int dt f(t)B(t),
\end{equation}
where $f(t)$ are step-like functions describing the periodic inversion of the NV spin for the pulse sequence under consideration~\cite{Cywinski2008,DeSousa2009}, and $B(t)$ includes the random magnetic fields due to the spin bath surrounding the NV center. 

Considering we have a large number of impurity spins with different couplings to the NV center, we assume that the random fluctuating B-field due to the spin bath has a Gaussian distribution (with zero mean), simply due to the Central Limit Theorem~\cite{DeLange2010}. Invoking properties of Gaussian noise, Eqn.\,\ref{eq:ProbrandomB} simplifies to
\begin{equation} 
p(\tau)=\frac{1}{2}\left( 1+e^{-\langle \phi(\tau)^2\rangle/2}\right)=\frac{1}{2}\left( 1+e^{-\chi(\tau)} \right),
\label{eq:NV0prob}
\end{equation}
where
\begin{equation}
\chi(\tau)=\left( \frac{g\mu_B}{\hbar} \right)^2 \int \frac{d\omega}{2\pi} \frac{S_B(\omega)}{\omega^2} F(\omega \tau).
\end{equation}
Following the convention established by Ref.~\cite{Cywinski2008} and others, 
\begin{equation}
S_B=\int dt e^{i\omega t} \langle B(t) B(0) \rangle,
\end{equation}
is the classical B-field spectral noise density and
\begin{equation}
F(\omega\tau)=\frac{\omega^2}{2}|\tilde{f}(\omega)|^2
\end{equation}
the filter function of the pulse sequence under consideration. The properties of the random B-field imposed here make it equivalent to treating it as an Orenstein-Uhlenbeck stochastic process~\cite{Klauder1962,DeSousa2009,DeLange2010,Witzel2014}.

We now look into the microscopic origin of the fluctuating B-field. For a bath spin Larmor precessing at $\omega_L$, the effective B-field at the central spin is given by $B_1 \cos{(\omega_L t)}$. Assuming the coherent spin precession decays at a timescale characterized by the bath correlation time $\tau_c$, the B-field at the central spin can be approximated by $B(t) = B_1 \cos{(\omega_L t)}e^{-t/\tau_c}$. Thus the bath spectral density becomes
\begin{eqnarray}\nonumber
S_B [\omega] &=& \int dt e^{i\omega t} \langle B_1^2 \rangle \cos{(\omega_L t)}e^{-t/\tau_c}\\
&=& \left\{ \frac{\tau_c  \langle B_1^2 \rangle}{1+ (\omega-\omega_L)^2\tau_c^2}  + \frac{\tau_c  \langle B_1^2\rangle}{1+ (\omega+\omega_L)^2\tau_c^2}  \right\}.
\end{eqnarray}
In the short time (high frequency) limit, the bath spectral density function can be simplified to 
\begin{equation}
S_B[\omega] \approx \langle B_1^2 \rangle \tau_c \frac{2}{1+(\delta\tau_c)^2}.
\end{equation}

For brevity and consistency with the main text, we define $\Delta^2_{\rm single}\equiv (g\mu_B/\hbar)^2\langle B_1^2 \rangle$. For an NV free induction decay (FID) measurement, the filter function is given by $F(\omega \tau) = 2 \sin^2{(\omega \tau/2)}$~\cite{Cywinski2008}, leading to 
\begin{eqnarray}\nonumber
\chi_\text{FID}(\tau) &=& \frac{2}{\pi}\Delta^2_{\rm single} \tau_c  \int \frac{d\omega}{\omega^2}  \frac{1}{1+(\delta\tau_c)^2} \sin^2{(\omega \tau/2)} \\
&=&\Delta^2_{\rm single} \tau_c^2 \left( \frac{t}{\tau_c}-1+e^{-t/\tau_c} \right). 
\end{eqnarray}

For a spin echo measurement, the filter function is given by $F(\omega \tau) = 8 \sin^4{(\omega \tau/4)}$~\cite{Cywinski2008}, giving us 
\begin{eqnarray}\nonumber
\chi_\text{SE}(\tau) &=& \frac{8}{\pi}\Delta^2_{\rm single} \tau_c  \int \frac{d\omega}{\omega^2}  \frac{1}{1+(\delta\tau_c)^2} \sin^4{(\omega \tau/4)} \\
&=&\Delta^2_{\rm single} \tau_c^2 \left( \frac{t}{\tau_c}-3-e^{-t/\tau_c}+4e^{-t/2\tau_c} \right). 
\end{eqnarray}

For short times, i.e. $t\ll\tau_c$ (but $\Delta_{\rm single }\tau_c\gg1$), $\chi$ for the two measurements simplifies to 
\begin{eqnarray}\nonumber
\chi_{FID}(\tau) &\approx&  \frac{\Delta^2_{\rm single}}{2} t^2,\\
\chi_{SE}(\tau) &\approx&  \frac{\Delta^2_{\rm single}}{12 \tau_c} t^3. 
\end{eqnarray}
Substituting this in Eqn. \ref{eq:NV0prob}, we get
\begin{eqnarray}
p_\text{FID}^{\rm single}(t) &=&\frac{1}{2}\left( 1+e^{-(t/T_{2,\rm single}^*)^2}\right),\\
p_\text{SE}^{\rm single}(t) &=& \frac{1}{2}\left( 1+e^{-(t/T_{2,\rm single})^3}\right),
\end{eqnarray}
where
\begin{equation}
T_{2,\rm single}^*=\left[ \frac{2 }{\Delta^2_{\rm single}} \right]^{1/2},
\label{eq:T2starsingleNV}
\end{equation}
\begin{equation}
T_{2,\rm single}=\left[ \frac{12 \tau_{c,\rm single}}{\Delta^2_{\rm single}} \right]^{1/3}.
\label{eq:T2singleNV}
\end{equation}
Here, we have replaced $\tau_c$ with $\tau_{c,\rm single}$ to be consistent with the main text. Finally, comparing this with the phenomenological quantum model of the spin bath dynamics developed in this work, we get
\begin{eqnarray}
\Delta^2_{\rm single} &=& \sum_i (C_{||}^{\text{NV},i}/2)^2,\label{eq:bathB0}\\
\frac{1}{\tau_{c,\rm single}} &=&\sum_{{\rm P1}, \{i,j\}}R^{ij}_{\rm flip}\label{eq:Hij}.
\label{eq:bathtauc}
\end{eqnarray}
The time dependencies for various dynamical decoupling protocols become different if we average this signal over several NV centers, each evolving under a random spin bath distribution, as in experiments. For dipolar interactions, the probability density function (PDF) for $\Delta_\text{single}$ has the form~\cite{Dobrovitski2008},

\begin{equation}
P(\Delta_\text{single})= \frac{\Delta_\text{ens}}{\Delta_\text{single}^2}\sqrt{\frac{2}{\pi}} e^{-\Delta_\text{ens}^2/2\Delta_\text{single}^2}.
\label{eq:PDFdelta}
\end{equation}

Here, $\Delta_\text{ens}$ is the average coupling strength of the NV to the spin bath within the NV ensemble. Integrating $p_\text{FID}^\text{single}(t)$ over the distribution of $\Delta_\text{single}$, we arrive at the ensemble averaged probability,
\begin{equation}\label{eq:FIDensm}
p_\text{FID}^\text{ens}(t) =\frac{1}{2}\left[ 1+e^{-(t/T_{2,\text{ens}}^*)}\right],
\end{equation}
where $T_{2,\text{ens}}^*=1/\Delta_\text{ens}$. The ensemble averaged FID signal thus exhibits simple exponential decay ($p=1$) in agreement with our experimental results (see Fig.\,\ref{fig:fig2}b inset).

We now perform a similar analysis to get an expression for the ensemble-averaged NV decoherence time $T_{2,\text{ens}}$. In this case, we need to take into account the nitrogen electronic spin bath dynamics. We classically model the bath correlation time as the time taken for a stochastic process (here several bath bath spin flip-flops) to reach a certain threshold, thus $\tau_{c,\rm single}$ can be seen as the equivalent of first passage time. The probability distribution function (PDF) of $\tau_c$ here is assumed to be a simplified Gaussian PDF given by
\begin{equation}
P(\tau_\text{c,single})=\sqrt{\frac{\lambda}{2 \pi \tau_\text{c,single}^3}} e^{-\lambda(\tau_\text{c,single}-\tau_\text{c,ens})^2/2\tau_\text{c,single}\tau_\text{c,ens}^2}.
\label{eq:PDFtauc}
\end{equation}
Here, $\lambda$ is an overall fitting parameter, and $\tau_\text{c,ens}$ is the ensemble mean of the distribution. As shown in Fig.\,\ref{fig:fig4}, the above PDFs model the numerical distribution of $\Delta$ and $\tau_c$ obtained from the quantum model. Finally, integrating over the distributions of $\Delta_\text{single}$ and $\tau_\text{c,single}$, we get for the spin echo signal 
\begin{equation}
p_\text{SE}^\text{ens}(t) \approx\frac{1}{2}\left[ 1+e^{-(t/T_{2,\text{ens}})^{3/2}}\right],
\label{eq:SEensm}
\end{equation}
where $T_{2,\text{ens}}=(2\tau_\text{c,ens}/\Delta_\text{ens}^2)^{1/3}$, and the ensemble-averaged decay exhibits non-integer decay with $p \approx 3/2$. 

\section{Numerical Simulations}
In order to simulate a random mixed electronic spin bath and its dynamics, we start with a diamond lattice putting an NV at origin, and pick random lattice sites for nitrogen P1 centers with the right concentration. Current simulations include $0.1 - 1000$\,ppm of nitrogen spins. We evaluate the dipole-dipole interaction between the NV and P1 spins to obtain $\Delta_\text{single}$ given by Eqn. \ref{eq:bathB0}. To obtain $\tau_\text{c,single}$, we sum all the P1 spin pair interactions according to Eqn.\,\ref{eq:Hij}. Finally, we extract the ensemble averaged values $\Delta_\text{ens}$ and $\tau_\text{c,ens}$ from the distribution generated from $\sim 10^4$ bath realizations (see Fig.\,\ref{fig:fig4} in main text for $[\text N=100\,$ppm). When numerically estimating $\tau_\text{c,single}$, we ignore spin bath pairs that interact weakly with the NV, leading to motional narrowing, as discussed in literature~\cite{DeSousa2009,DeLange2010}.


\bibliographystyle{unsrt}
\bibliography{references}

\end{document}